\documentclass{ws-procs10x7}
\begin{document}
%
\title{
\flushright{UCHEP-04-04} \\
\center{{\boldmath $B$} decays to open and hidden charm at Belle}}
\author{A. Drutskoy}
\address{Physics Department, University of Cincinnati, \\
345 College Court, Cincinnati, OH 45221, USA}
\twocolumn[\maketitle\abstract{
The recent Belle collaboration measurements 
of $B$ decays to open and hidden charm are discussed.
Color-suppressed decay branching fractions are measured
with an improved accuracy. The branching fractions of the 
$\bar{B}^0 \to D_s^+ K^-$ and $\bar{B}^0 \to D_s^- \pi^+$
decays, measured with improved accuracy, and  
$\bar{B}^0 \to D_{sJ}^+ K^-$ and $\bar{B}^0 \to D_{sJ}^- \pi^+$
decays, measured for the first time,
are compared.
The two-body invariant masses of the three-body
$B^0 \to D^{(*)0} \pi^+ \pi^-$ and $B^0 \to J/\Psi \pi^+ \pi^-$
decays are studied.}]
\section{Results}
\subsection{Color-suppressed $\bar{B}^0 \to D^0 \pi^0 (/\omega/\eta)$ 
and $\bar{B}^0 \to D^{(*)0} \eta'$ decays}\label{subsec:prod}
The weak decays $\bar{B}^0 \to D^{(*)0} h^0$, 
where $h^0$ represents a light neutral
meson, are usually described by a ``color-suppressed'' diagram as
shown in Fig.1.
\vspace{-45pt}
\begin{figure}[h!]
\epsfxsize110pt
\figurebox{110pt}{150pt}{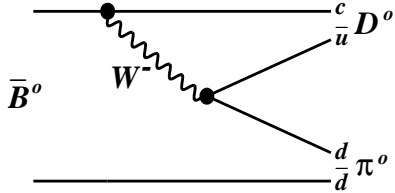}
\caption{Color-suppressed diagram for the \newline decay
$\bar{B}^0 \to D^{(*)0} \pi^0$.}
\label{fig:radk}
\end{figure}

Within ``na\"ive'' factorization models \cite{facta,factb},
the color-matching requirement \linebreak
leads to branching fractions in the range 
\mbox{(0.3-1.7)$\times 10^{-4}$}. 
However previous measurements by the CLEO~\cite{dpa} , Belle~\cite{dpb} 
and BaBar~\cite{dpc} collaborations
were substantially shifted to the larger 
values (2-4)$\times 10^{-4}$.  
The new Belle measurements are based on a larger data sample of 140 fb$^{-1}$.
This corresponds to a seven-fold increase
over the previous Belle measurement and almost twice 
that of the earlier BaBar measurement.
With increased statistics a more detailed study of 
continuum and $B\bar{B}$ related backgrounds was done,
reducing systematic uncertainties.

The Belle measurements of these branching fractions 
for each decay mode is shown in Table~1.

\begin{table}[h!]
\caption{$B$ branching fractions for color-suppressed decays.}\label{tab:smtab}
\begin{tabular}{|l|c|} 
\hline 
\raisebox{0pt}[10pt][3pt]{Decay mode} &
\raisebox{0pt}[10pt][3pt]{Br. Fraction $(\times 10^{-4})$} \\
\hline
\raisebox{0pt}[10pt][3pt]{$\bar{B}^0 \to D^0 \pi^0$} & 
\raisebox{0pt}[10pt][3pt]{2.31 $\pm$ 0.12 $\pm$ 0.23} \\
\raisebox{0pt}[10pt][3pt]{$\bar{B}^0 \to D^0 \eta (\gamma\gamma)$} & 
\raisebox{0pt}[10pt][3pt]{1.77 $\pm$ 0.18 $\pm$ 0.20} \\
\raisebox{0pt}[10pt][3pt]{$\bar{B}^0 \to D^0 \eta (\pi^0\pi\pi)$} & 
\raisebox{0pt}[10pt][3pt]{1.89 $\pm$ 0.29 $\pm$ 0.20} \\
\raisebox{0pt}[10pt][3pt]{$\bar{B}^0 \to D^0 \eta$} & 
\raisebox{0pt}[10pt][3pt]{1.83 $\pm$ 0.15 $\pm$ 0.27} \\
\raisebox{0pt}[10pt][3pt]{$\bar{B}^0 \to D^0 \omega (\pi^0\pi\pi)$} & 
\raisebox{0pt}[10pt][3pt]{2.25 $\pm$ 0.21 $\pm$ 0.28} \\
\raisebox{0pt}[10pt][3pt]{$\bar{B}^0 \to D^0 \eta'$} & 
\raisebox{0pt}[10pt][3pt]{1.17 $\pm$ 0.20 $^{+0.10}_{-0.14}$} \\
\raisebox{0pt}[10pt][3pt]{$\bar{B}^0 \to D^{*0} \eta'$} & 
\raisebox{0pt}[10pt][3pt]{1.23 $\pm$ 0.34 $\pm$ 0.21} \\
\hline
\end{tabular}
\end{table}

These branching fractions are about one-two standard deviations
lower than the BaBar measurements but higher than early 
predictions within factorization models. The latter discrepancy 
may be explained
by additional contributions from final state rescattering 
or non-factorisable diagrams.  

\subsection{Improved measurement of $\bar{B}^0 \to D_s^+ K^-$
and $\bar{B}^0 \to D_s^- \pi^+$ decays and
first study of $\bar{B}^0 \to D_{sJ}^+ K^-$ and 
$\bar{B}^0 \to D_{sJ}^- \pi^+$ decays}\label{subsec:wpp}

\begin{figure*}
\vspace{-10pt}
\epsfxsize120pt
\figurebox{120pt}{160pt}{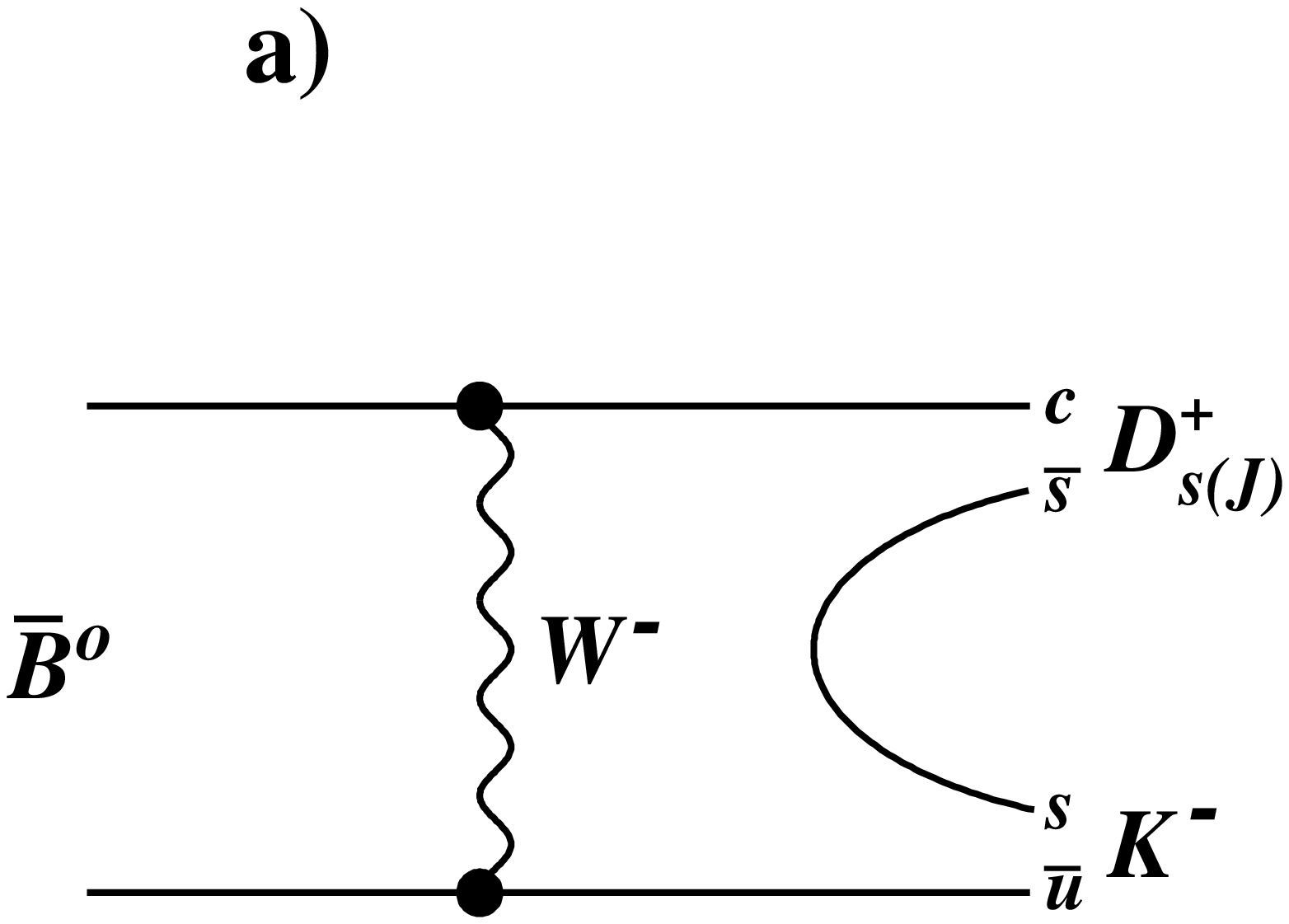}
\epsfxsize120pt
\figurebox{120pt}{160pt}{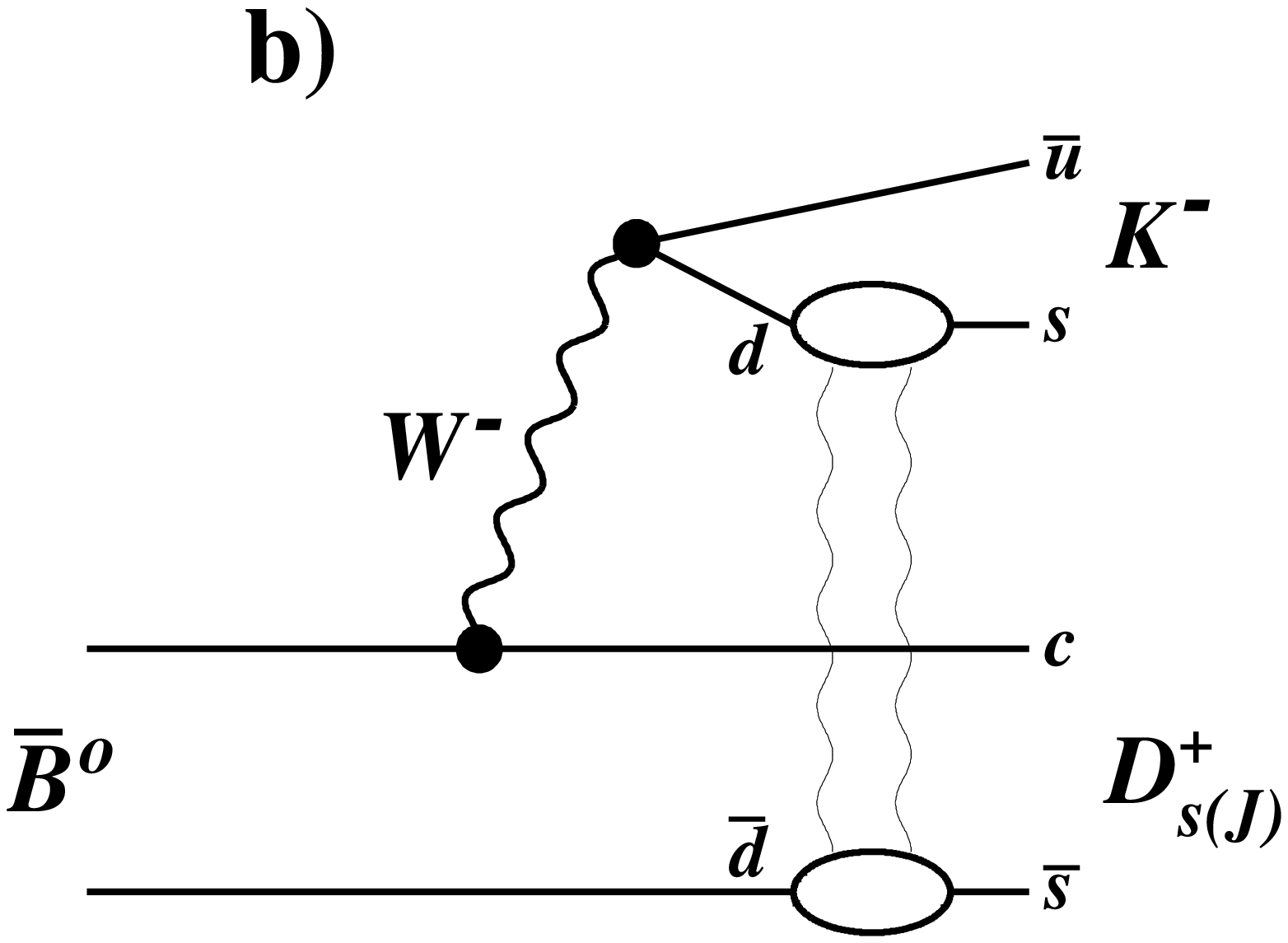}
\epsfxsize120pt
\figurebox{120pt}{160pt}{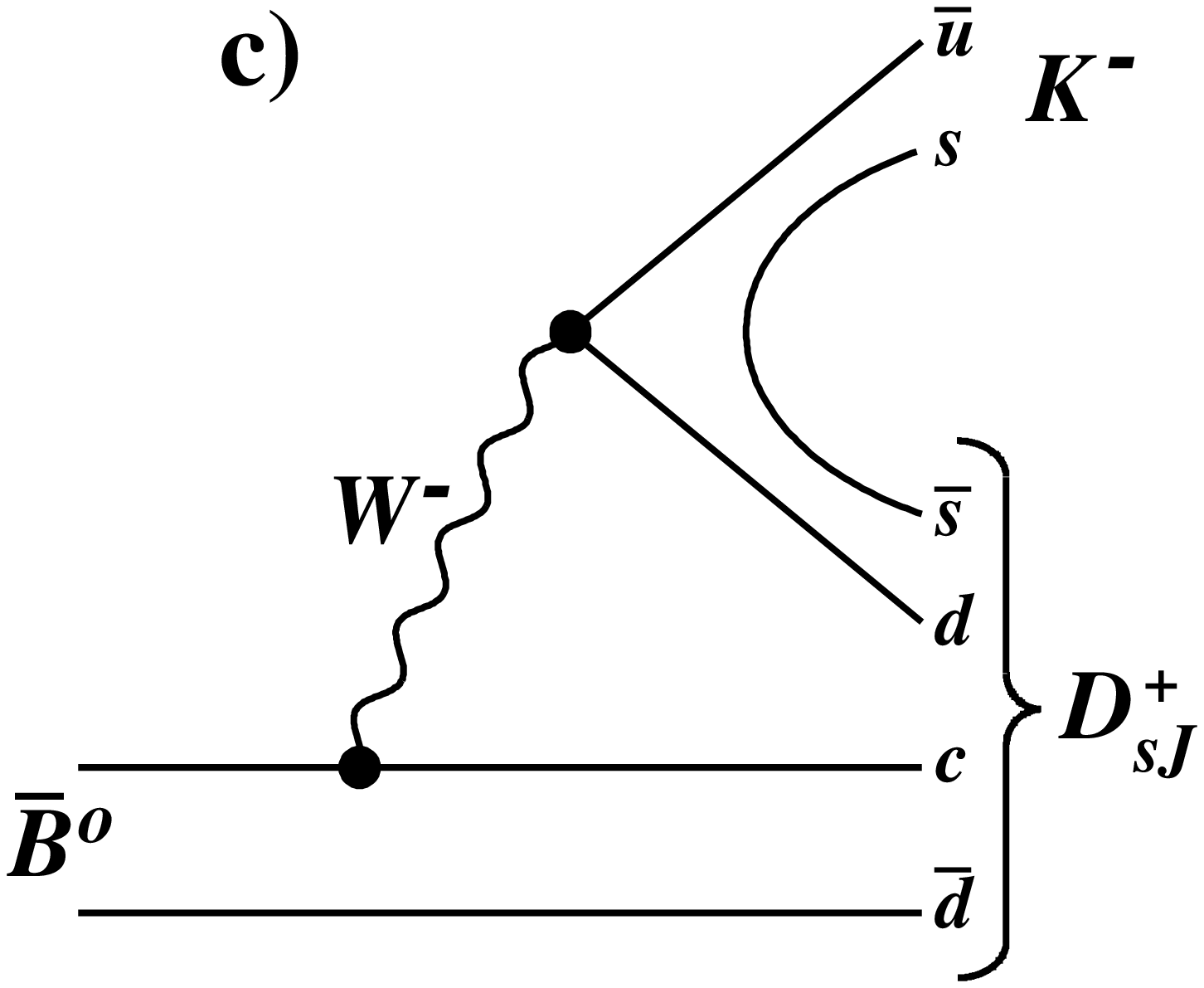}
\caption{Diagrams describing 
$\ensuremath{\overline{B}{}^0} \to D_{sJ}^+ K^-$ decay.\label{fig:radish}}
\vspace{-10pt}
\end{figure*}

The decays $\bar{B}^0 \to D_{s(J)}^+ K^-$ are of special interest because 
the quark content of the initial $\bar{B}^0$ meson ($b\bar{d}$)
is completely different from that of the $D_{s(J)}^+ K^-$ 
final state ($cs\bar{s}\bar{u}$),
indicating an unusual configuration with both initial quarks 
involved in the weak decay.
Branching fractions with the $D_s^+$ meson
${\cal B}(\bar{B}^0 \to D_s^+ K^-) = (4.6^{+1.2}_{-1.1} \pm 1.3)\cdot 10^{-5}$ and
$(3.2 \pm 1.0 \pm 1.0)\cdot 10^{-5}$
have been measured by the Belle~\cite{belc} and BaBar~\cite{babc}
collaborations, respectively.
Predictions of this branching fraction have been obtained
assuming a dominant contribution from a PQCD factorization $W$ 
exchange process~\cite{kteoa,kteob} (Fig.~2a)
or, alternatively, from final state 
interactions~\cite{kteoc,kteod} (Fig.~2b),
and range from a few units of $10^{-6}$ to $10^{-4}$. 
If the $D_{sJ}$ mesons have a four-quark component
then the tree diagram with $s\bar{s}$ pair 
creation (shown in Fig.~2c) may also contribute.

The decay mode $\bar{B}^0 \to D_{s(J)}^- \pi^+$ can be described by
a ``$b$ to $u$'' tree diagram.
Within the factorization approach\cite{ratio} the branching fraction ratio
$R_{\pi^+/D^+} = {\cal B}(\bar{B}^0 \to D_s^- \pi^+) / 
{\cal B}(\bar{B}^0 \to D_s^- D^+)$
is predicted to be $(0.424 \pm 0.041) \cdot |V_{ub}/V_{cb}|^2$ 
and can be used to obtain the ratio of Cabbibo-Kobayashi-Maskawa 
matrix elements $|V_{ub}/V_{cb}|$. 

The $B$ decay channels with the pseudoscalar $D_s$ meson
were studied using 253 fb$^{-1}$ of data
(275$\times$10$^6$$B\bar{B}$ pairs), whereas channels with the $D_{sJ}$
were studied using 140 fb$^{-1}$ of data 
(152$\times$10$^6$ $B\bar{B}$ pairs).
$D_s^+$ mesons are reconstructed in the $\phi \pi^+$, $K^{*0} K^+$
and $K_S^0 K^+$ decay channels. $D_{sJ}$ mesons
are reconstructed in the $D_{sJ}^\ast(2317)^+ \to D_s^+ \pi^0$ and
$D_{sJ}(2460)^+ \to D_s^+ \gamma$ decay modes.

The $D_s$ mass distributions in the $B$ signal
region are shown in Fig. 3. The points are 
experimental data and the curves display the fit results.
In addition to clear signals at the $D_s^+$ mass in Fig. 3,
the $D^+$ mass peak ($D^+$ decays in the same final states)
is also seen.

\vspace{2pt}
\begin{figure}[h!]
\epsfxsize190pt
\figurebox{190pt}{235pt}{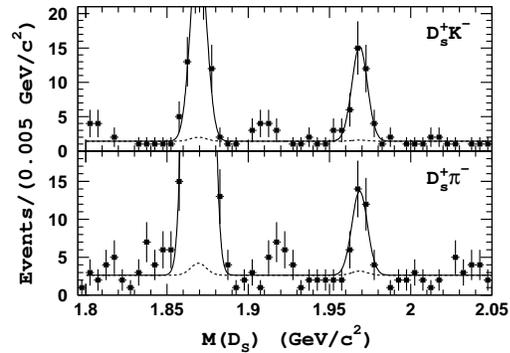}
\caption{$M(D_s)$ spectra for $\bar{B}^0 \to D_s^+ K^-$ (top)
and $\bar{B}^0 \to D_s^- \pi^+$ (bottom) in the $B$ signal
region.}
\label{fig:radk}
\end{figure}

The branching fractions obtained from the fits are 
${\cal B}(\bar{B}^0 \to D_s^+ K^-) = (2.93 \pm 0.55 \pm 0.79)\cdot 10^{-5}$
and ${\cal B}(\bar{B}^0 \to D_s^- \pi^+) = (1.94 \pm 0.47 \pm 0.52)\cdot 10^{-5}$. 
These agree within errors with the previous measurements.

The $\Delta M(D_{sJ})$ distributions in the $B$ signal region
for various $D_{sJ}^+ K^-$ and $D_{sJ}^- \pi^+$ combinations
are shown in Fig.~4. A clear $\bar{B}^0 \to D_{sJ}^\ast(2317)^+ K^-$ 
signal is observed;
no significant signals are observed in the remaining modes.

\begin{figure}[h!]
\epsfxsize190pt
\figurebox{190pt}{225pt}{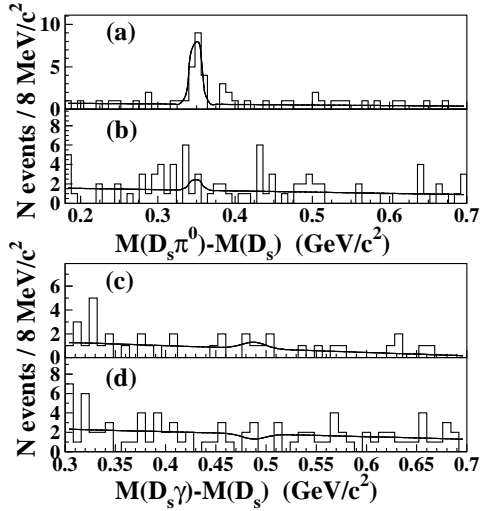}
\caption{The $\Delta M(D_{sJ})$ distributions in the $B$ signal region
for the (a) $D_{sJ}^\ast(2317)^+ K^-$, (b) $D_{sJ}^\ast(2317)^- \pi^+$,
(c) $D_{sJ}(2460)^+ K^-$ and (d) $D_{sJ}(2460)^- \pi^+$ combinations
are shown.}
\label{fig:radk}
\end{figure}

For the 
$\bar{B}^0 \to D_{sJ}^\ast(2317)^+ K^-$ decay,
the product branching fraction is measured to be
${\cal B}(\bar{B}^0 \rightarrow D_{sJ}^\ast(2317)^+ K^-) \times
{\cal B}(D_{sJ}^\ast(2317)^+ \rightarrow D_s^+ \pi^0) =$
\mbox{$(5.3^{+1.5}_{-1.3} \pm 0.7 \pm 1.4) \cdot 10^{-5}$}.
Recent measurements imply that the $D_{sJ}^\ast(2317)^+ \to$ \linebreak
$D_s^+ \pi^0$ channel is dominant and the \linebreak
$D_{sJ}(2460)^+ \to D_s^+ \gamma$ fraction is around 30$\%$.
Taking into account these approximate values, we can conclude that
$\mathcal{B}(\bar{B}^0 \to D_{sJ}^\ast(2317)^+ K^-)$ 
is of the same order of magnitude as $\mathcal{B}(\bar{B}^0 \to D_s^+ K^-)$,
but at least a factor two larger than
the $\bar{B}^0 \to D_{sJ}(2460)^+ K^-$ branching fraction,
in contrast to the na\"{\i}ve expectation that decays with
the same spin-doublet 
$D_{sJ}^\ast(2317)^+$ and $D_{sJ}(2460)^+$ mesons would have similar rates.
It is interesting to mention, that the ratio of 
$\mathcal{B}(B \to D_{sJ}^\ast(2317)^+ D)$ and
$\mathcal{B}(B \to D_s^+ D)$ decay branching fractions~\cite{dsd}
is approximately 1/10,
indicating to a different behaviours of the $B \to D_s K$
and $B \to D_s D$ processes~\cite{dskd}.

\subsection{Study of $B^0 \to D^{(*)0} \pi^+ \pi^-$ decays}

The three-body $B^0 \to D^{(*)0} \pi^+ \pi^-$ decays
were studied with 140 fb$^{-1}$ of data. 
These processes can provide important information
about intermediate two-body resonances.
The $D^0 \pi^+$ and $D^{*0} \pi^+$ mass distributions
are shown in Figs. 5,6.
Points are experimental data, the hatched histogram is obtained
using sideband events and the open histogram is MC 
simulation with all intermediate resonances included.
The masses and widths of narrow ($\leq$100 MeV/c$^2$)
$D^{**}$ resonances observed are listed in Table 2.
These values are in a good agreement with predictions
obtained within potential models~\cite{dmod}.

\vspace{-6pt}
\begin{figure}[h!]
\epsfxsize147pt
\figurebox{147pt}{190pt}{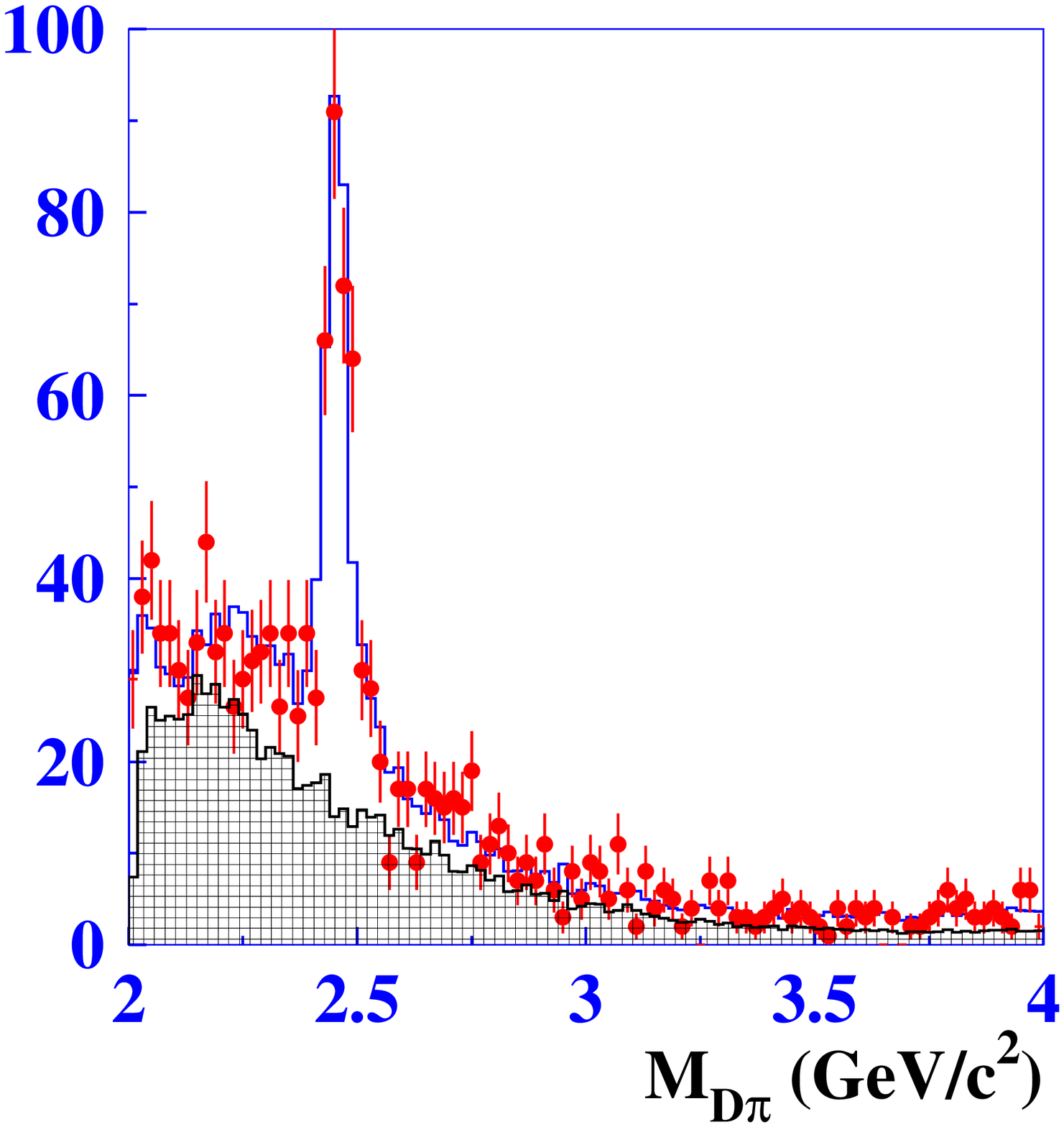}
\caption{
The $D^0 \pi^+$ mass distribution.}
\label{fig:radk}
\end{figure}
\vspace{-12pt}
\begin{figure}[h!]
\epsfxsize147pt
\figurebox{147pt}{190pt}{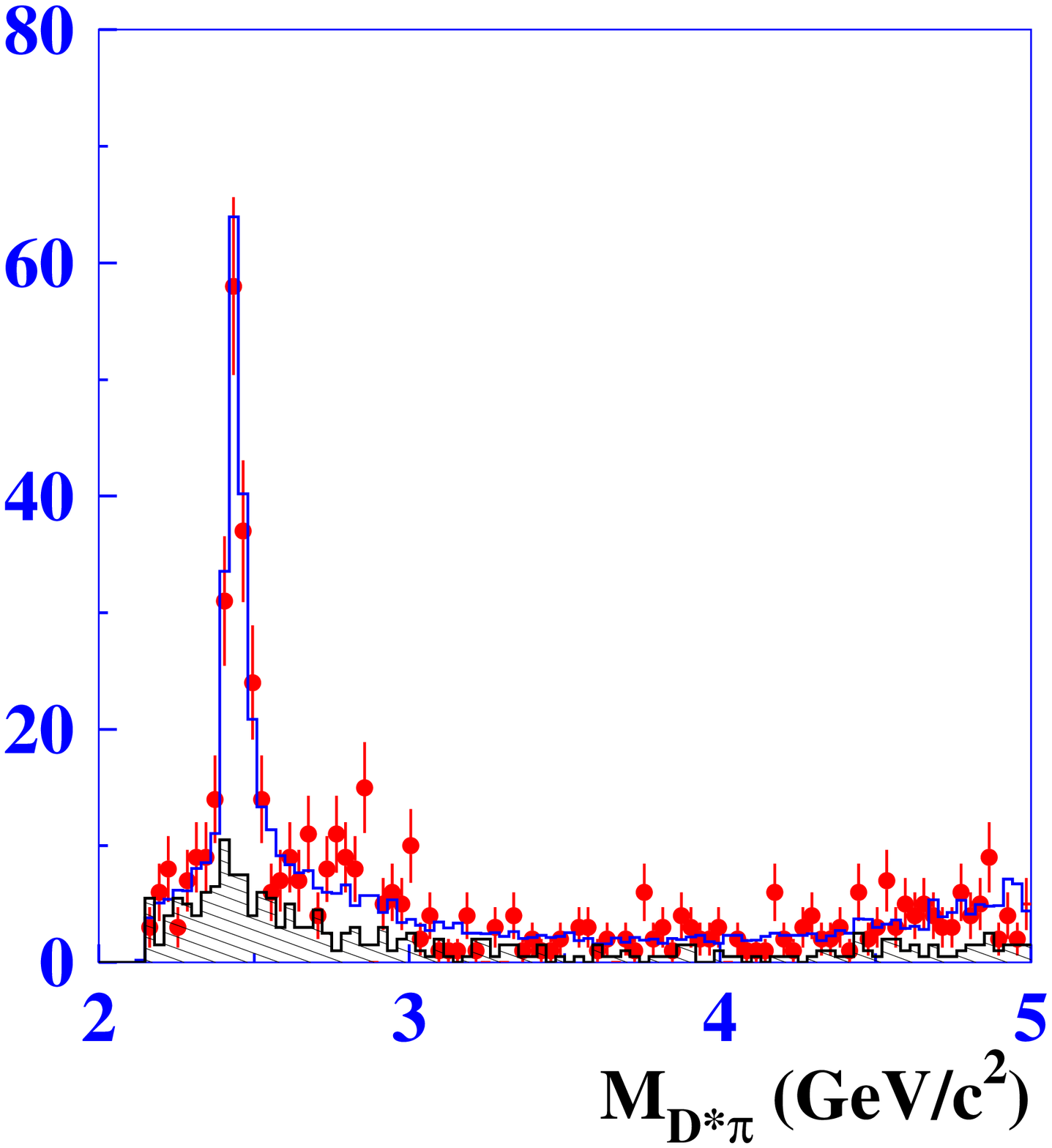}
\caption{
The $D^{*0} \pi^+$ mass distribution.}
\label{fig:radk}
\end{figure}
\vspace{-1pt}

\begin{table}[h!]
\caption{Masses and widths of narrow $D^{**}$ resonances.}\label{tab:smtab}
\begin{tabular}{|l|c|} 
\hline 
\raisebox{0pt}[10pt][3pt]{Parameter} &
\raisebox{0pt}[10pt][3pt]{Value, MeV/c$^2$} \\
\hline
\raisebox{0pt}[10pt][3pt]{$M(D_2^{*+})$} & 
\raisebox{0pt}[10pt][3pt]{2459.5 $\pm$ 2.3 $\pm$ 0.7 $^{+4.9}_{-0.5}$} \\
\hline
\raisebox{0pt}[10pt][3pt]{$\Gamma (D_2^{*+})$} & 
\raisebox{0pt}[10pt][3pt]{48.9 $\pm$ 5.4 $\pm$ 4.2 $\pm$ 1.9} \\
\hline
\raisebox{0pt}[10pt][3pt]{$M(D_1^+)$} & 
\raisebox{0pt}[10pt][3pt]{2428.2 $\pm$ 2.9 $\pm$ 1.6 $\pm$ 0.6} \\
\hline
\raisebox{0pt}[10pt][3pt]{$\Gamma(D_1^+)$} & 
\raisebox{0pt}[10pt][3pt]{34.9 $\pm$ 6.6 $^{+4.1}_{-0.9}$ $\pm$ 4.1} \\
\hline
\end{tabular}
\end{table}

The quasi-two-body $\bar{B}^0 \to D^{**+} \pi^-$ final state
branching fractions were measured
and compared to results obtained by the Belle collaboration in the studies
of the charged $B^+ \to D^{(*)-} \pi^+ \pi^+$ decay modes~\cite{belch}.
The branching fractions obtained for the narrow resonances are
similar for the neutral and charged $B$ decays.

\subsection{Study of $B^0 \to J/\Psi \pi^+ \pi^-$ decays}

The decay $B^0 \to J/\Psi \rho^0$ is governed by the
$b \to c\bar{c}d$ transition and can exhibit a $CP$-violating
asymmetry. In contrast to the $b \to c\bar{c}s$ transition,
the $b \to c\bar{c}d$ process has substantial contributions
from both the tree and penguin amplitudes, which could lead
to different $CP$ asymmetries of these processes.
Thus, $B^0 \to J/\Psi \rho^0$ decays play an important role
in probing non-tree diagram contributions. 

The resonant structure of the $\pi^+ \pi^-$ invariant mass spectrum from
$B^0 \to J/\Psi \pi^+ \pi^-$ decays was studied using 140 fb$^{-1}$
of data. Five types of events are considered in the fit
shown in Fig. 7: (i) $B^0 \to J/\Psi \rho^0$; 
(ii) $B^0 \to J/\Psi f_2$; (iii) $B^0 \to J/\Psi \pi^+ \pi^-$ (non-resonant);
(iv) $B^0 \to J/\Psi K_S^0$ (background) and (v) combinatorial background.
The branching fractions obtained are
${\cal B} (B^0 \to J/\Psi \rho^0) = (2.8 \pm 0.3 \pm 0.3) \times 10^{-5}$ and
${\cal B} (B^0 \to J/\Psi f_2) = (9.8 \pm 3.9 \pm 2.0) \times 10^{-6}$.
The statistical significance of the latter is 2.9$\sigma$, and an upper
limit is also set: ${\cal B} (B^0 \to J/\Psi f_2) < 1.5 \times 10^{-5}$
at 90$\%$ C.L. An upper limit is also set on the non-resonant 
channel:
${\cal B} (B^0 \to J/\Psi (\pi^+ \pi^-)_{non-res.}) < 1.0 \times 10^{-5}$
at 90$\%$ C.L.

\begin{figure}[h!]
\epsfxsize160pt
\figurebox{160pt}{200pt}{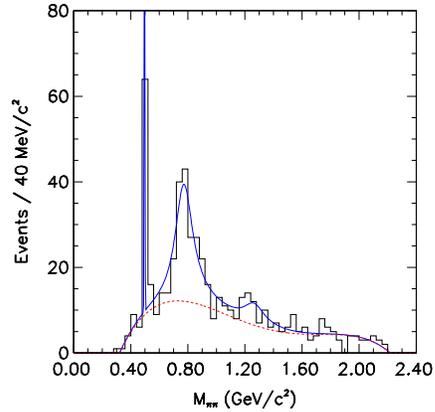}
\caption{
The distribution of $M_{\pi^+ \pi^-}$ for $B^0 \to J/\Psi \pi^+ \pi^-$
decay. Points are the experimental data. The solid line shows the result 
of fit with the $K_S^0$, $\rho^0$, $f_2$ and background contributions
included, the dashed line is for the background contribution only.}
\label{fig:radk}
\end{figure}


\begin{thebibliography}{99}
\bibitem{facta}
M. Neubert and A.A. Petrov, {\it Phys. Lett.} B {\bf 519}, 50 (2001).
\bibitem{factb}
A. Deandrea and A.D. Polosa, {\it Eur. Phys. Jour.} C {\bf 22}, 677 (2002).
\bibitem{dpa}
CLEO Collaboration, T.E. Coan {\it et al.}, {\it Phys. Rev. Lett.} {\bf 88}, 062001 (2002).
\bibitem{dpb}
Belle Collaboration, K. Abe {\it et al.}, {\it Phys. Rev. Lett.} {\bf 88}, 052002 (2002).
\bibitem{dpc}
BaBar Collaboration, B. Aubert {\it et al.}, {\it Phys. Rev.} D {\bf 69}, 032004 (2004).
\bibitem{belc}
Belle Collaboration, P. Krokovny {\it et al.}, {\it Phys. Rev. Lett.} {\bf 89},
231804 (2002). 
\bibitem{babc}
BaBar Collaboration, B. Aubert {\it et al.}, {\it Phys. Rev. Lett.} {\bf 90},
181803 (2003).
\bibitem{kteoa} 
D. Du, L. Guo, D.-X. Zhang, {\it Phys. Lett.} B {\bf 406},
110 (1997).
\bibitem{kteob}
C.D. Lu, hep-ph/0305061.
\bibitem{kteoc}
C.-K. Chua, W.-S. Hou, K.-C. Yang, {\it Phys. Rev.} D {\bf 65},
096007 (2002).
\bibitem{kteod}
B. Blok, M. Gronau, J.L. Rosner, {\it Phys. Rev. Lett.} {\bf 78},
3999 (1997).
\bibitem{ratio}
C.S. Kim, Y. Kwon, J. Lee, W. Namgung, {\it Phys. Rev.} D {\bf 63},
094506 (2001).
\bibitem{dsd}
Belle Collaboration, P. Krokovny {\it et al.}, {\it Phys. Rev. Lett.} {\bf 91},
262002 (2003).
\bibitem{dskd} C.-H. Chen, H.-n Li, {\it Phys. Rev.} D {\bf 69},
054002 (2004).
\bibitem{dmod}
S. Godfrey and R. Kokoski, {\it Phys. Rev.} D {\bf 43},
1679 (1991).
\bibitem{belch}
Belle Collaboration, K. Abe {\it et al.}, {\it Phys. Rev.} D {\bf 69},
112002 (2004).
\end{thebibliography}
\end{document}